\documentclass[final,3p,times,twocolumn]{elsarticle}

\usepackage{amssymb}

\usepackage{amsmath,bm}
\usepackage{bm}
\usepackage{xcolor}
\usepackage{hyperref}

\definecolor{urlcolor}{HTML}{799B03}
\hypersetup{pdfstartview=FitH,  linkcolor=linkcolor,urlcolor=urlcolor, colorlinks=true}

\begin{document}

\begin{frontmatter}



\title{Twisted-photons Spectral-Angular Distribution Emitted by Relativistic Electrons at Axial Channeling}



\author[1]{K.B. Korotchenko\corref{cor1}}\ead{korotchenko@tpu.ru}

\author[2]{Y.P. Kunashenko\fnref{fn2}}\ead{iurii.p.kunashenko@tusur.ru}

\cortext[cor1]{Corresponding author}
\fntext[fn2]{These author contributed equally to this work.}

\affiliation[1]{organization={Division for Experimental Physics,
National Research Tomsk Polytechnic University},
            addressline={30 Lenina Ave.},
            city={Tomsk},
            postcode={634050},
            country={Russia}}

\affiliation[2]{organization={Department of mathematics,
Tomsk State University of Control Systems and Radioelectronics},
            addressline={40 Lenina Ave.},
            city={Tomsk},
            postcode={634050},
            country={Russia}}

\begin{abstract}
 Within the framework of QED, spectral-angular distribution of the twisted photons emitted by relativistic electrons during axial channeling were investigated.
\end{abstract}


\end{frontmatter}


\section{Introduction}\label{sec1}

 In our previous work, \cite{KorotchenkoRPC} we investigated the distribution of twisted photons emitted by relativistic electrons during axial channeling. We have developed a theory of the radiation of twisted photons by axially channeled electrons within the framework of quantum electrodynamics. This theory allowed us to draw for the first time the angular distribution of twisted photons emitted by channeled electrons. Here we continue our study of the radiation of twisted photons. Our new goal is the spectral-angular distributions of the photons emitted by axially channeled relativistic electrons.

 Twisted photons are not plane waves, but superpositions of plane waves with a certain projection of orbital angular momentum onto its propagation direction \cite{Knyazev1}. Twisted photons are characterized by longitudinal momentum, modulus of transverse momentum, and projection of total angular momentum onto longitudinal momentum.

 The theory of twisted photons within the framework of quantum electrodynamics was developed by Serbo et al. Based on this theory, were developed methods for obtaining beams of twisted photons and possible experiments with them were studied \cite{Jentschura, Jentschura1, Scholz}. The generation of twisted photons opens up new possibilities for studying photonuclear reactions and offers new tools in nuclear physics. Twisted radiation has found numerous applications in both classical and quantum condensed matter optics, high-energy physics, optics, etc. (see for example \cite{Vieira, Knyazev1, Afanasev, Abramochkin, Torres, Andrews} and references therein).

 In Bogdanov et al. (\cite{Bogdanov1, Bogdanov2, Bogdanov3, Bogdanov4, Bogdanov5, Bogdanov6} a theory of emission of twisted photons by relativistic charged particles under various conditions was developed. These works are based on the semi-classical approach developed by Baier et al. \cite{Baier}. The main disadvantage of these works is that they consider the radiation of photons in a strictly forward direction. However, photons can be emitted at any angle relative to the particle’s velocity \cite{KorotchenkoRPC}.

It is well known that channeling radiation is much more intense than ordinary bremsstrahlung. At present, this effect allows the creation of intense radiation sources \cite{Uberall, Kumakh, Baryshevsky, Bazylev, Baier, Akhiezer, Kimball, Kumakhov, Artru}.

 We hope that the channeling of electrons in crystals will also increase the intensity of radiation of twisted photons compared to that in an amorphous target.

 Axially channeled electrons have orbital momentum according to both the classical and quantum models. An electron can transfer part of its angular momentum to a photon during emission, causing the photon to become twisted.

\section{ Twisted-photon radiation probability}\label{sec3}

 Our study of the spectral distribution of twisted photons  (TW-photons) emitted by relativistic channeled electrons is based on the results of \cite{KorotchenkoRPC}.

 From here on we use the notation from the article \cite{KorotchenkoRPC}. According to \cite{KorotchenkoRPC}, the probability of twisted photon radiation from a channeled electron (TWcr-photons) has the form
\begin{equation}
    dW_{fi} = \frac{2\pi}{\hbar}|M_{fi}|^2\delta(\mathcal{E}_i - \mathcal{E}_f - \hbar\omega) \frac{\omega}{c^2}\frac{d\omega d\theta_\kappa}{2\pi^2}\frac{L dp_z}{2\pi\hbar}
    \: . \label{eq1}
\end{equation}

 The matrix element of TWcr-photon emission $M_{fi}$ is equal to
\begin{align}
  M_{fi}^{cr} &= \bm{i}^{-m}\frac{c\hbar}{2}m_{fi}^{cr}\sqrt{\frac{\alpha}{\pi LR}\sin\theta_\kappa}\: , \label{eq2}\\
  m_{fi}^{cr} &= \sum_{m_s=-1}^1\int \bm{\alpha}_{fi} d_{m_s\Lambda}^{1}(\theta_\kappa){\bm{\varepsilon}}_{CR}^{m_s} e^{\bm{i}(m-m_s)\varphi_\kappa} e^{\bm{i}\kappa_Z(\varphi_\kappa)Z} d\varphi_\kappa \nonumber\\
  & \frac{1}{L}\int e^{-\bm{i}\frac{\Delta p_{fi,Z}}{\hbar}Z}dZ\: .\nonumber
\end{align}
 Here $d_{m_s\Lambda}^{1}(\theta_\kappa)$ is the the Wigner (small) d-matrix, $m_s = \pm 1$ is the projection of the spin onto a given axis, $\theta_\kappa$ and $\varphi_\kappa$ are the TW-photon ``internal'' polar and azimuthal angles, $\Lambda$ is the TW-photon helicity, ${\bm{\varepsilon}}_{CR}^{m_s}$ is the polarization vector of a TW-photon in the coordinate system associated with the photon \cite{KorotchenkoRPC}.

 The vector $\bm{\alpha}_{fi}$, can be written as follows \cite{Korotchenko}
\begin{equation}
 {\bm{\alpha}}_{fi} = \Big(\frac{\Omega_{fi}}{c}\kappa_Y(\varphi_\kappa),\: \frac{\Omega_{fi}}{c}\kappa_X(\varphi_\kappa),\: \beta \kappa_X(\varphi_\kappa) \kappa_Y(\varphi_\kappa)\Big) \langle XY\rangle_{fi}\: , \label{eq3}
\end{equation}
where
\begin{align}
    \kappa_X & = \kappa_z\sin\Theta\cos\Phi + \varkappa(\cos\Theta-1)\cos\Phi\sin\varphi_\kappa\sin\Phi + \nonumber\\
    & \varkappa\cos\varphi_\kappa(\cos\Theta\cos^2\Phi + \sin^2\Phi)\: , \nonumber\\
    \kappa_Y & = \kappa_z\sin\Theta\sin\Phi + \varkappa(\cos\Theta-1)\cos\Phi\cos\varphi_\kappa\sin\Phi + \nonumber\\
    &  \varkappa\sin\varphi_\kappa(\cos\Theta\sin^2\Phi + \cos^2\Phi)\: , \nonumber\\
    \kappa_Z & = \kappa_z\cos\Theta - \varkappa\cos(\varphi_\kappa - \Phi)\sin\Theta\: . \label{eq04}
\end{align}
 Here we used the relations for the TW-photon from \cite{KorotchenkoRPC, Knyazev1}
\begin{equation}
    \hbar\omega = c\hbar\sqrt{\varkappa^2 + \kappa_z^2}, \: \: \: \tan\theta_\kappa = \frac{\varkappa}{\kappa_z}\: .  \label{eq05}
\end{equation}
 The other notations in (\ref{eq3}):
 $\beta = v_{\|}/c$ ($v_{\|}$ is the longitudinal velocity of the channeled electron), $\hbar\Omega_ {fi} = \mathcal{E}_i - \mathcal{E}_f$ and $\mathcal{E}_ {i (f)}$ are the energies of the $i (f)$-th transverse quantum state, and $\langle XY\rangle_{fi}$ equals \cite{Korotchenko}
\begin{equation}
 \langle XY\rangle_{fi} =
  C_{i,i_n}^{m_i,n_i}C_{f,f_n}^{m_f,n_f}\frac{(-1)^{m_f+m_i+n_f+n_i}a_p^2}{16 \pi^2 (m_f-m_i)(n_f-n_i)}\: . \label{eq06}
\end{equation}
 where $C_{i,i_n}^{m_i,n_i}$ are the Fourier components of the channeled electron transverse wave function, $m_i$ and $n_i$ are the Fourier components numbers for the $i$-th energy band of the channeled electron transverse motion, and $a_p$ is the lattice constant \cite{KorotchenkoRPC}.

\section{TWcr-photon spectral-angular distributions}\label{sec:5}

 As in \cite{KorotchenkoRPC, Korotchenko}, we sum the probabilities of emission of TWcr-photons by the photon helicities.

 Similarly \cite{Baier, Kumakh, Baryshevsky, Bazylev, Akhiezer, Kimball}, we find in \cite{KorotchenkoRPC}, that the energy conservation law for TWcr-photons is $\hbar\omega \approx \hbar(\beta c\Delta + \Omega_{fi})$ where $\Delta = \Delta p_{fi}/\hbar$ and $\Delta p_{fi} = p_{z_i} - p_{z_f}$ (see (\ref{eq2})).

 After integrating the probability (\ref{eq1}) over the longitudinal momentum of the electron $dp_z$ (i.e. $\Delta$), taking into account the $\delta$-function, we obtain\footnote{When integrating over $\Delta$, the $\delta$-function ``removes'' integration over $\Delta$ and replaces $\Delta$ with $\omega/(c \beta)$.}
\begin{equation}
 \frac{dW_{fi}}{d\theta_\kappa d\omega} = \frac{\alpha}{16\hbar\pi^3 c \beta}\sin\theta_\kappa \langle XY\rangle_{fi}^2 \Big(\frac{\omega}{c}\Big)^4 |m_{fi}^{cr}(\Theta, \Phi, \theta_\kappa, \omega)|^2\: .
\label{eq07}
\end{equation}

 Compared to \cite{Bogdanov1, Bogdanov2, Bogdanov3, Bogdanov4, Bogdanov5, Bogdanov6} this probability additionally depends on the angles $\Theta$ and $\Phi$ describing the direction of the TWcr-photon  vector relative to the crystal reference frame. Therefore, the angles $\Theta$ and $\Phi$ are additional parameters to the usual variables of the TWcr-photon wave function.

 Considering that for a total angular momentum (TAM) $m \geq 3$ of a TWcr-photon the probability of its emission is symmetrical with respect to the angle $\Phi$ (see \cite{KorotchenkoRPC}), we can average the probability (\ref{eq07}) over this angle. Then for the ``intensity'' $dI_{fi} = \hbar\omega dW_{fi}$ we can obtain
\begin{equation}
 \frac{dI_{fi}}{d\theta_\kappa d\omega} = \frac{\alpha}{16\pi^3\beta}\sin\theta_\kappa \langle XY\rangle_{fi}^2 \Big(\frac{\omega}{c}\Big)^5 |\langle m_{fi}^{cr}(\Theta, \theta_\kappa, \omega)\rangle_\Phi|^2\: .
\label{eq08}
\end{equation}

 The resulting equation (\ref{eq08}) still depends parametrically on the emission angle $\Theta$ of the TWcr-photon, so we call it the spectral-angular distribution. As in \cite{KorotchenkoRPC}, we perform the calculation for $\omega \gg \Omega_{fi}$ (i.e. for X-ray photons TWcr).

\begin{figure}[h]
\centering\noindent
\includegraphics[width=8cm]{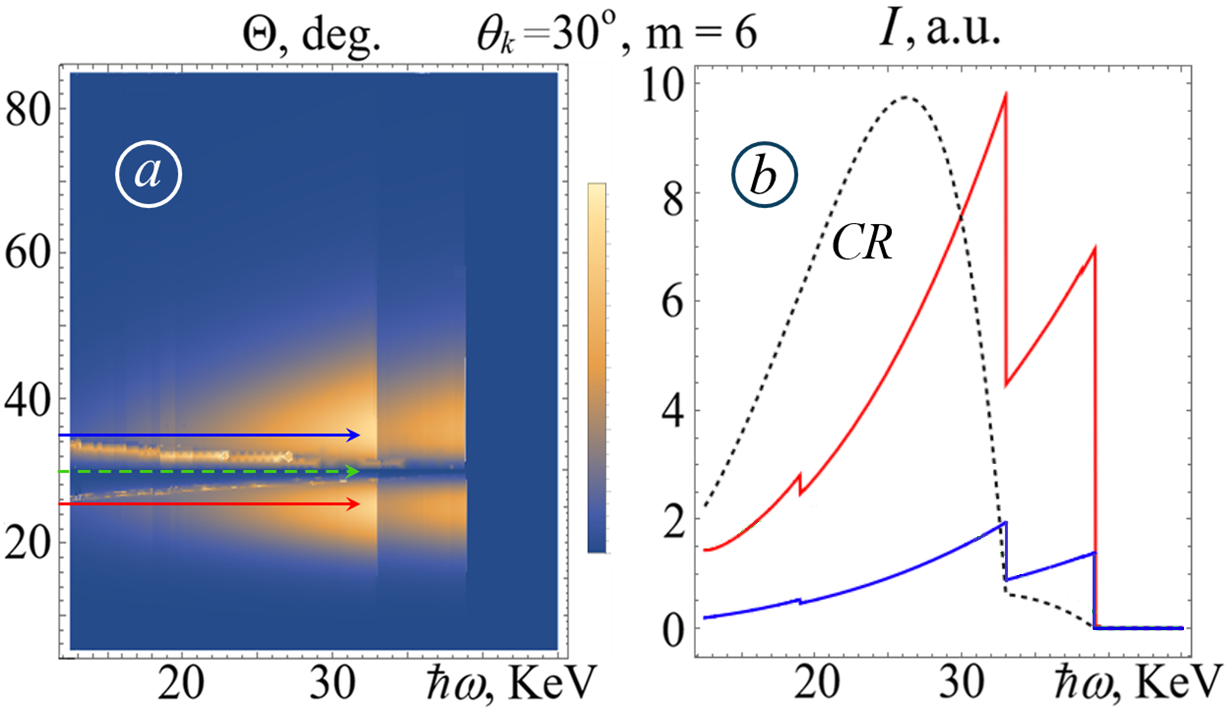}
\caption{Spectral-angular distribution of X-ray TWcr-photons with TAM $z$-projection $m = 6$ and ``internal'' angle $\theta_\kappa = 30^{\circ}$. a) Dependence of the ``intensity'' $I$ on TWcr-photon energy $\hbar\omega$ and emission angle $\Theta$ (density plot). The green dashed arrow indicates the angle $\theta_\kappa$. b) Dependence of the ``intensity'' (arbitrary units used) on TWcr-photon energy $\hbar\omega$ for two emission angles, blue line - $\Theta \simeq 35.7^{\circ}$, red line $\Theta \simeq 25.4^{\circ}$. The black dashed line indicates the spectral distribution CR from axially channeled electrons (see \cite{Korotchenko}).}\label{fig:1}
\end{figure}

 Figure~\ref{fig:1} shows the calculated spectral-angular distributions of the ``intensity''
\begin{equation}
 I = \sum_{fi}\frac{dI_{fi}}{d\theta_\kappa d\omega}P_i(\theta_0)
\label{eq09}
\end{equation}
 TWcr-photons with $z$-projection of the total angular momentum (TAM) $m = 6$ and ``internal'' angle $\theta_\kappa = 30^{\circ}$. Here and below, the calculations were performed for electrons with an energy of $10$ MeV channeled along the $\langle 100\rangle$ axes of the Si crystal (as in \cite{KorotchenkoRPC, Korotchenko}).

 We summed the intensity $dI_{fi}$ over all possible transitions of a channeled electron between transverse energy bands and considered the initial population of the bands $P_i(\theta_0)$. Here $\theta_0$ is the angle of the electron's longitudinal momentum relative to the crystal axes (for more details, see \cite{KorotchenkoRPC, Korotchenko}). The angle $\theta_0 = 2\theta_C/15$ is used in our calculations, where $\theta_C$ is the critical angle of the channeling.

 The initial population of the energy levels of channelled electrons is the probability of electron capture into channelled states, a well-known characteristic has been studied in detail (see, e.g., \cite{Korotchenko} and references therein).

\begin{figure}[h]
\centering\noindent
\includegraphics[width=8cm]{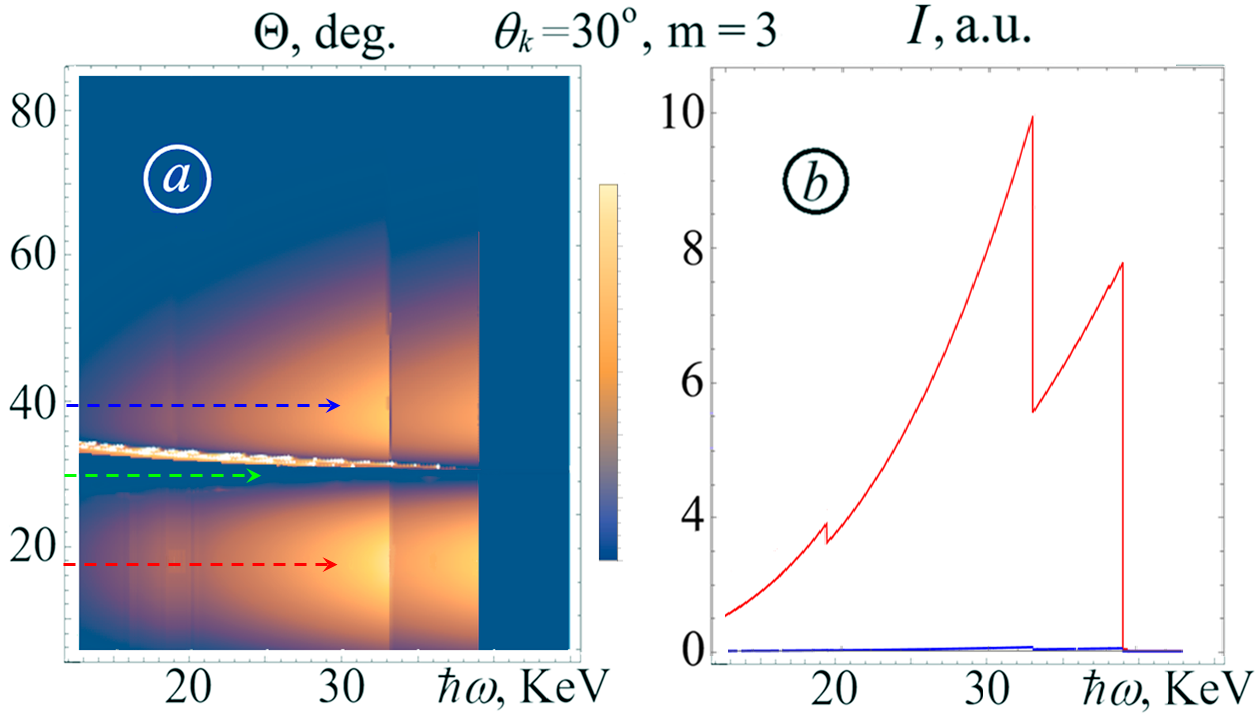}
\caption{The same as Fig.\ref{fig:1} but for $m = 3$ and  $\theta_\kappa = 30^{\circ}$;  $\Theta \simeq 35.5^{\circ}$ (blue line) and  $\Theta \simeq 24.9^{\circ}$ (red line). The green dashed arrow indicates the angle $\theta_\kappa$.}\label{fig:2}
\end{figure}

\begin{figure}[h]
\centering\noindent
\includegraphics[width=8cm]{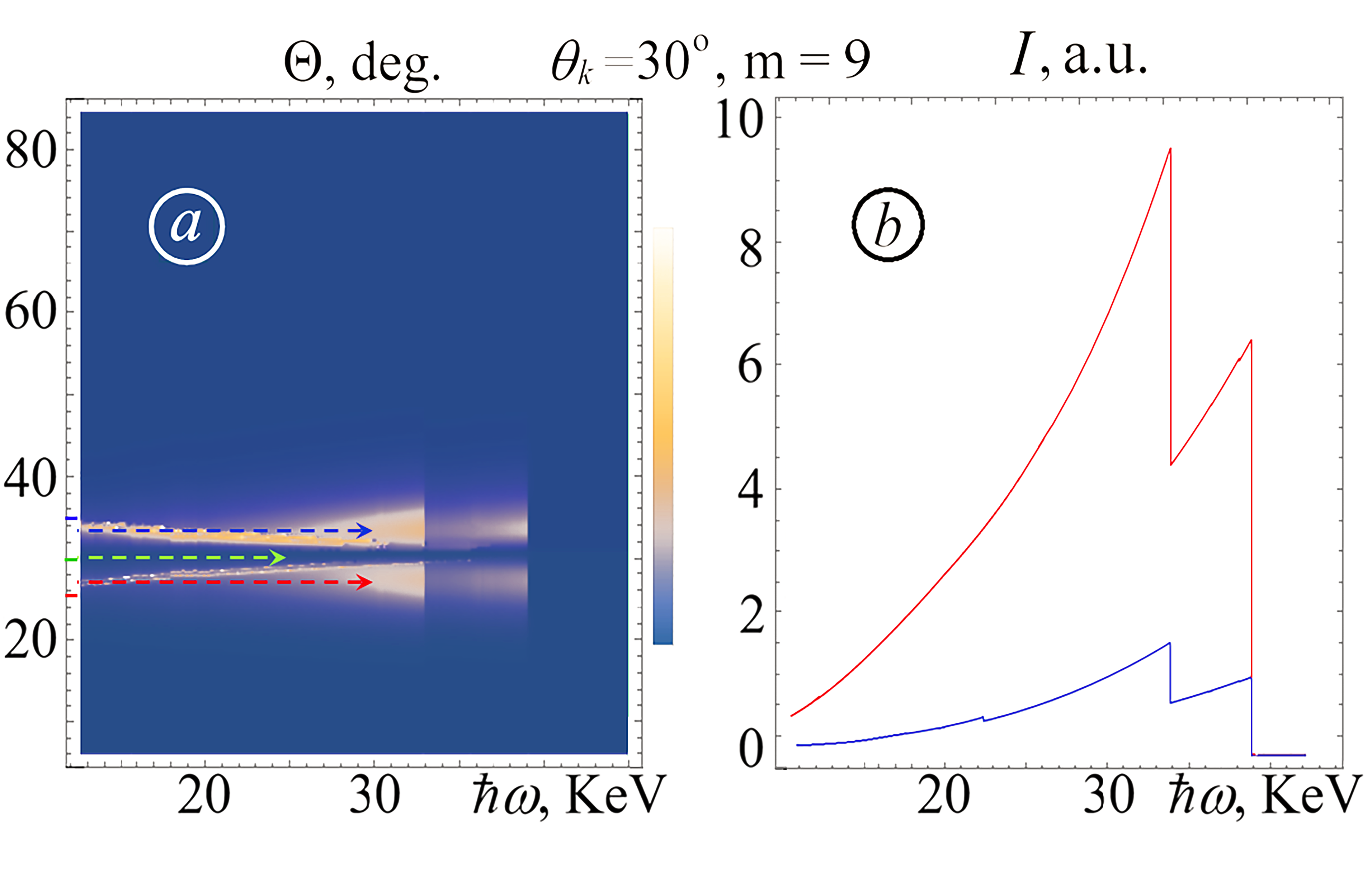}
\caption{The same as Fig.\ref{fig:1} but for $m = 9$, $\theta_\kappa = 30^{\circ}$;  $\Theta \simeq 33.4^{\circ}$ (blue line) and  $\Theta \simeq 26.4^{\circ}$ (red line). The green dashed arrow indicates the angle $\theta_\kappa$.}\label{fig:3}
\end{figure}

 Fig.\ref{fig:1} (and further Figs.\ref{fig:2}-\ref{fig:3}) clearly shows that the spectral angular distribution of  TWcr-photons has two maxima. The angular positions of the maxima are indicated by color arrows. The density plots only shows the position of the spectral angular distribution of TWcr-photons in the ``angular-energy plane'' and does not allow for a numerical comparison of ``intensities'' for different angles and energies.

 Therefore, in Fig.\ref{fig:1}b (and further Figs.\ref{fig:2}b-\ref{fig:3}b) we plot the dependence of ``intensity'' on the TWcr-photon energy for two emission angles $\Theta \simeq 35.7^{\circ}$ and $\Theta \simeq 25.4^{\circ}$. The color of the lines is the same as for arrows.

 As is well known, the intensity of Channeling Radiation (CR) from relativistic electrons has abrupt drop points at certain photon frequencies  (see e.g. \cite{Korotchenko}). It can be seen that the ``intensity'' $I$ in Fig.\ref{fig:1}b (and further) also drops sharply at certain photon frequencies. This emphasizes that we are dealing with CR of twisted photons by channeled electron.

 In Fig.\ref{fig:1}a (and further Figs.\ref{fig:2}a-\ref{fig:3}a), the ``bright'' regions are triangular. The higher the photon energy, the brighter the triangle (the higher the ``intensity''). At a certain energy, the triangle's width decreases, and it darkens. This energy coincides with the energy at which the sharp drop in ``intensity'' is observed (Fig.\ref{fig:1}b-\ref{fig:3}b)).

 Figs.\ref{fig:2} and \ref{fig:3} show the calculated spectral-angular distributions of the ``intensity'' $I$ of X-ray TWcr-photons with $z$-projection of TAM $m = 3$ and $m = 9$ for the same ``internal'' angle $\theta_\kappa = 30^{\circ}$.

\begin{figure}[h]
\centering\noindent
\includegraphics[width=8cm]{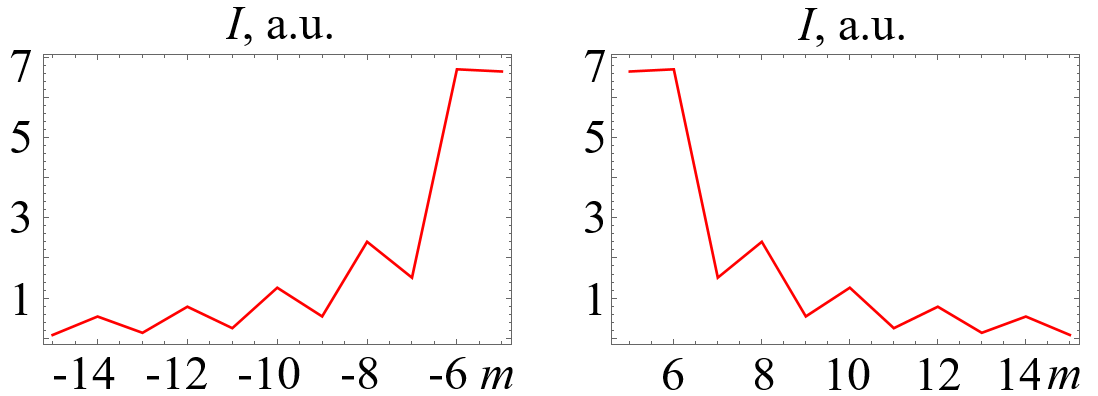}
\caption{Spectral distribution of TAM $z$-projection $m = 5...15$ and $m = -5...-15$ for X-ray TWcr-photons with ``internal'' angle $\theta_\kappa = 60^{\circ}$.}\label{fig:4}
\end{figure}

 Fig.\ref{fig:4} shows the ``spectrum'' of TAM $m$ in the ranges $m = 5...15$ and $m = -5...-15$.

 The most surprising result presented in Figs.\ref{fig:1} - \ref{fig:3} is that the value ``internal'' angle $\theta_\kappa$ defines the azimuthal angle $\Theta = \Theta_O$, near which X-ray TWcr-photons are emitted.

 The intensity has two maxima one at angles slightly smaller $\Theta = \theta_\kappa$ and second at slightly larger $\Theta = \Theta_O$.

 To verify this, calculations of the spectral-angular distributions of X-ray TWcr-photons with an ``internal'' angle $\theta_\kappa = 60^{\circ}$ were performed. The results are shown in Fig.\ref{fig:5}.
\begin{figure}[h]
\centering\noindent
\includegraphics[width=8cm]{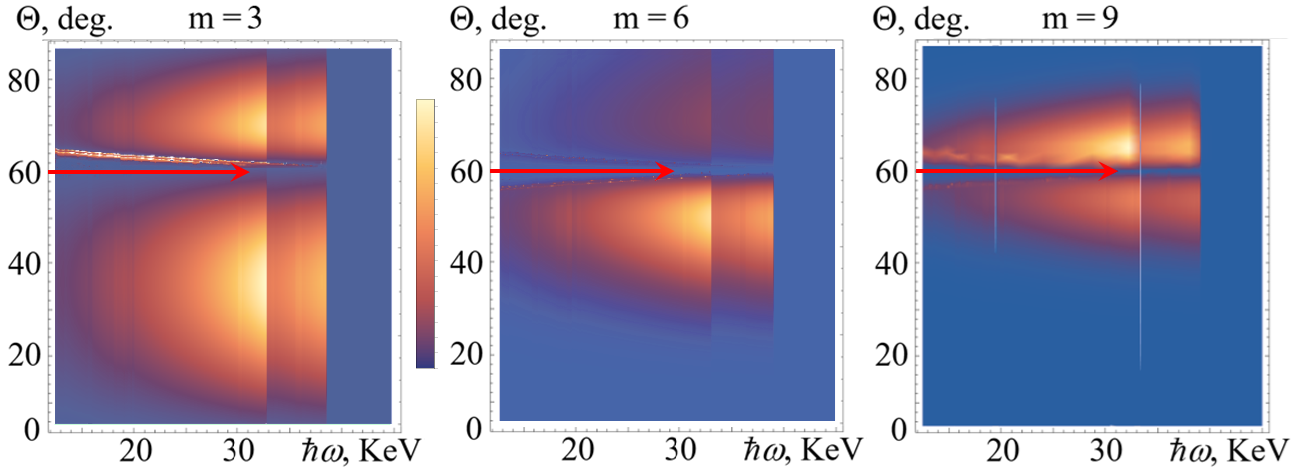}
\caption{Spectral-angular distribution of X-ray TWcr-photons with TAM $z$-projection $m = 3, 6, 9$ and ``internal'' angle $\theta_\kappa = 60^{\circ}$.}\label{fig:5}
\end{figure}

 The ``intensity'' $I$ of TWcr-photons emitted at the angles $\Theta < \Theta_O$ is greater than the ``intensity'' of TWcr-photons emitted at the angles $\Theta > \Theta_O$.

 These results can be visualized using an intensity plot $I = I(\Theta,  \theta_\kappa)$ of TWcr-photons (see Fig.\ref{fig:5}).
\begin{figure}[h]
\centering\noindent
\includegraphics[width=8cm]{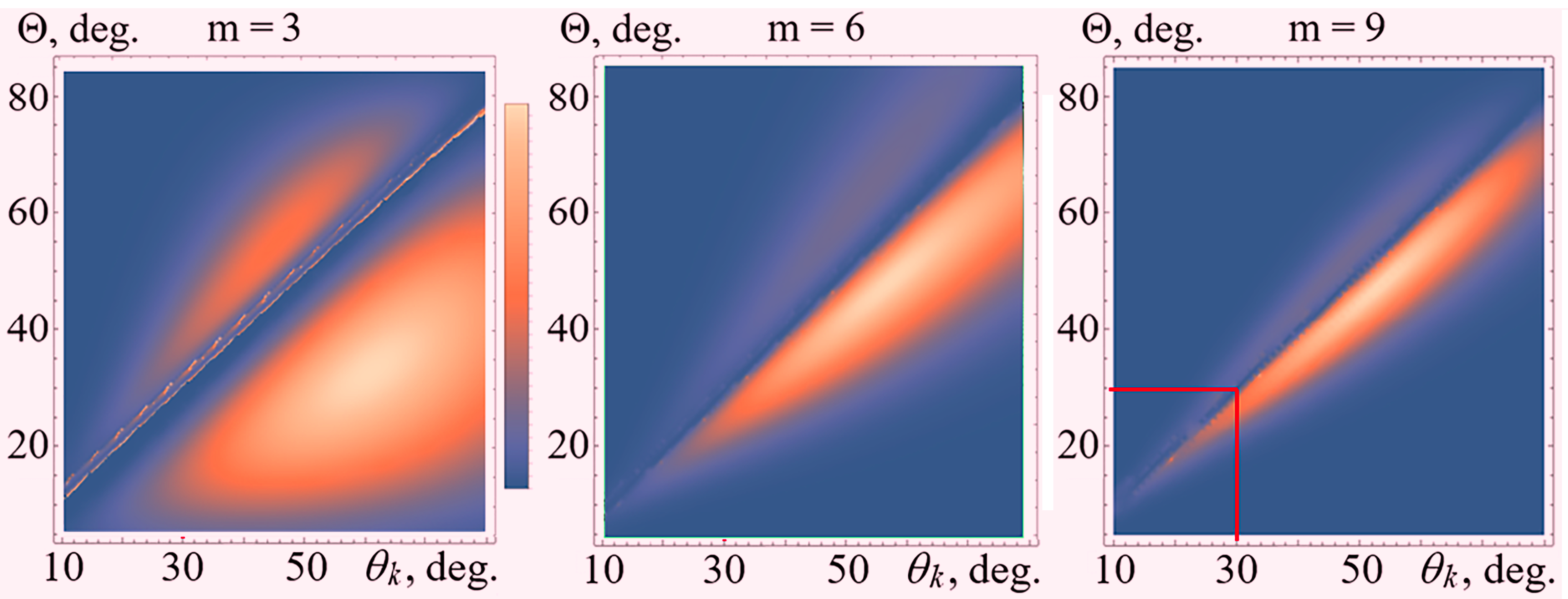}
\caption{Density plot for ``intensity'' $I = I(\Theta,  \theta_\kappa)$ of TWcr-photons with TAM $z$-projection $m = 3, 6, 9$ and energy $\hbar\omega = 30$KeV.}\label{fig:5}
\end{figure}

\section{Conclusion}\label{sec:6}

 Within the framework of QED, the spectral-angular distributions of twisted photons emitted by relativistic electrons during the axial channeling are investigated.

In summary, we note:
\begin{itemize}
  \item As a function of energy, the probability of TWcr-photon emission has sharp peaks at certain energies. The positions of these peaks are the same as those of the peaks for ordinary CR (see Fig.\ref{fig:1} - \ref{fig:3}).
  \item The distribution of TWcr-photon has a two maxima near $\Theta = \theta_\kappa$. The ``intensity'' $I$ (\ref{eq09}) of TWcr-photons emitted at angles $\Theta \leq \theta_\kappa$ is greater than at angles $\Theta \geq \theta_\kappa$.
\end{itemize}

 In this work, we investigated a TWcr-photon's emission during the axial channeling of electrons at a large angle to the crystal axis. The case of radiation at small angles will be considered in a separate article.

\end{document}